\documentclass[12pt]{article}
\usepackage{amsmath,epsfig,amssymb,eufrak,dsfont}
\usepackage{amscd,latexsym}
\usepackage{color,ccaption}

\date{}

\def\xlf{\raisebox{+0.2em}{\color{red}\boldmath{$\chi$}}\hspace{-0.2ex}\raisebox{-0.2em}{\color{green}L}
\hspace{-1.5ex}\raisebox{+0.14em}{\color{blue}F}\hspace{2mm}}

\def\lsi{\raise0.3ex\hbox{$<$\kern-0.75em\raise-1.1ex\hbox{$\sim$}}}
\def\gsi{\raise0.3ex\hbox{$>$\kern-0.75em\raise-1.1ex\hbox{$\sim$}}}

\newcommand{\kcpi}{\kappa_c^{\rm pion}}
\newcommand{\kcP}{\kappa_c^{~\!\!\!_{\rm PCAC}}}

\input macros.sty      % \eqalign, \meqalign
\input eqalign.sty     % \eqalign, \meqalign

\begin{document}

\begin{titlepage}

\title{
  {\vspace{-1.5cm} \normalsize
  \hfill \parbox{40mm}{DESY/05-139\\
                       SFB/CPP-05-35\\November 2005}}\\[10mm]
Parton Distribution Functions\\ with Twisted Mass Fermions}
\author{S.~Capitani$^{\, 1}$, K.~Jansen$^{\, 2}$, M.~Papinutto$^{\, 2}$\hspace{-0.1cm}
\footnote{Present address: INFN Sezione di Roma 3, Via della Vasca Navale 84, I-00146 Roma, ITALY}, \\
A.~Shindler$^{\, 2}$, C.~Urbach$^{\, 2,\,3}$\hspace{-0.1cm}
\footnote{Present address: Theoretical Physics Division, Dept. of Mathematical Sciences, 
University of Liverpool, Liverpool L69 3BX, UK} and I.~Wetzorke$^{\, 2}$\\ 
\\
   {\bf \xlf Collaboration}\\
\\
  {\small $^{1}$  Institut f\"ur Physik/Theoretische Physik}\\
  {\small Universit\"at Graz, A-8010 Graz, Austria} \\ \ \\
  {\small $^{2}$  John von Neumann-Institut f\"ur Computing NIC,} \\
  {\small         Platanenallee 6, D-15738 Zeuthen, Germany} \\ \ \\
  {\small $^{3}$ Institut f\"{u}r Theoretische Physik, Freie Universit\"{a}t Berlin,} \\
  {\small Arnimallee 14, D-14195 Berlin, Germany} \\ 
}
\maketitle

\begin{abstract}
We present a first Wilson twisted mass fermion calculation of the 
matrix element between pion states of the twist-2 operator, which is related to
the the lowest moment $\langle x\rangle$ of the valence quark parton
distribution function in a pion.  
Using Wilson twisted mass fermions in the quenched approximation
we demonstrate that $\langle x\rangle$ can be computed at small
pseudoscalar meson masses
down to values of order 250 MeV. We investigate the scaling behaviour of 
this physically important quantity by applying two definitions of the
critical mass and observe a scaling compatible with the expected O($a^2$)
behaviour in both cases. A combined continuum extrapolation allows 
to obtain reliable results for $\langle x\rangle$ at very small pseudoscalar meson
masses, which previously could not be explored by lattice QCD
simulations. 
\vspace{0.75cm}
\noindent
%%%{\it PACS:}  ; ; \\
%%%{\it Keywords:}  ; .
\end{abstract}

\end{titlepage}

\section{Introduction}
In lattice QCD the so-called chiral extrapolation 
of data obtained at pseudoscalar meson masses of about 
500 MeV to the physical pion mass value (140 MeV) is one of 
the main systematic uncertainties in today's lattice QCD calculations. 
As an example from our own work let us quote a recent paper \cite{Guagnelli:2004ga}
where the lowest moment $\langle x\rangle$ of the valence quark parton distribution
function in a pion has been calculated, fully controlling the continuum
limit, finite size effects \cite{Guagnelli:2004ww}
and the non-perturbative renormalization
\cite{Guagnelli:2003hw}. The only remaining uncertainty, besides the quenched
approximation, was the chiral extrapolation. 

In this letter we want to report on a first step towards eliminating also this 
systematic error by employing the twisted mass formulation of lattice QCD
\cite{Frezzotti:2000nk}
which allows to regulate unphysically small eigenvalues of the Wilson-Dirac 
operator 
and simultaneously achieves an O($a$)-improvement of physical
observables without the need of improvement coefficients \cite{Frezzotti:2003ni}
(automatic O($a$) improvement).  
With the example of $\langle x\rangle$ 
we will demonstrate that indeed this formulation of lattice QCD 
allows to bridge the gap between results at pseudoscalar meson mass values of 500 MeV
or larger -- as obtained in conventional simulations -- and the physical value
of the pion mass. Performing the simulations at a number of values of the 
gauge coupling $g_0^2=6/\beta$, the continuum limit of $\langle x\rangle$ 
is performed at values of the pseudoscalar meson mass small enough to
allow, at least in principle, the 
comparison of the numerical results with the predictions from chiral 
perturbation theory. 

Wilson twisted mass fermions have been employed already in a 
number of quenched simulations                   
\cite{Jansen:2003ir,Bietenholz:2004sa,Bietenholz:2004wv,Abdel-Rehim:2004gx,Abdel-Rehim:2005gz,
Jansen:2005gf,Jansen:2005kk,Abdel-Rehim:2005qv,Abdel-Rehim:2005yx}. See refs.
\cite{Frezzotti:2004pc,Shindler:2005vj} for recent reviews.

Also full QCD simulations using this approach have been performed and
proved to be very useful in studying the phase 
structure of lattice QCD with Wilson type fermions
\cite{Farchioni:2004us,Farchioni:2004ma,Farchioni:2004fs,Farchioni:2005tu}.
On the theoretical side, various studies were performed by means of the
Symanzik expansion and of chiral perturbation theory
\cite{Sharpe:1998xm,Munster:2004am,Scorzato:2004da,Sharpe:2004ps,Aoki:2004ta,Sharpe:2004ny,Frezzotti:2005gi}. 

In recent quenched simulations 
\cite{Abdel-Rehim:2005gz,Jansen:2005gf,Jansen:2005kk} it has been shown
that the Wilson twisted mass approach can be used to simulate small pseudoscalar meson
masses of order 250 MeV while keeping O($a^2$) cut-off effects
under control, when employing 
the definition of the critical mass derived from the vanishing of the 
PCAC quark mass \cite{Aoki:2004ta,Sharpe:2004ny,Frezzotti:2005gi}. 
These simulations were done for basic observables as 
extracted from 2-point correlation functions. In this letter we will extend
the investigation of Wilson twisted mass fermions also to
the physically important case of 3-point functions, in particular
for matrix elements related to moments of parton distribution functions
which are relevant in deep inelastic scattering. 

\section{Lattice action and operators}

\subsection{Wilson twisted mass fermions}

In this letter we will work on a lattice $L^3 \times T$ 
with Wilson twisted mass fermions \cite{Frezzotti:2000nk} which can be 
arranged to be O($a$) improved without employing specific improvement terms
\cite{Frezzotti:2003ni}. 
The action for a degenerate flavour doublet of twisted mass fermions can be written as
\begin{equation}
  \label{tmaction}
  S[U,\psi,\bar\psi] = a^4 \sum_x \bar\psi(x) ( D_W + m_0 + i \mu
\gamma_5\tau_3 ) \psi(x)\; ,
\end{equation}
where the Wilson-Dirac operator $D_{\rm W}$ is given by
\be
D_{\rm W} = \sum_{\mu=1}^4 \frac{1}{2} 
[ \gamma_\mu(\nabla_\mu^* + \nabla_\mu) - a \nabla_\mu^*\nabla_\mu]
\label{Dw}
\ee
$\nabla_\mu$ and $\nabla_\mu^*$ denote the usual forward
and backward derivatives, $m_0$ and $\mu$ denote the untwisted and twisted
bare quark masses.
We refer to refs.~\cite{Bietenholz:2004sa,Bietenholz:2004wv} for further
unexplained notations.
Here and in the following $\psi(x)$ indicates a flavour doublet of quarks.  

A key element in this twisted mass setup is the definition of the critical 
quark mass $m_c$, since, in order to obtain automatic O($a$) improvement, the target
continuum theory should have a vanishing untwisted quark mass.
In the present work, we will employ two 
definitions for the critical mass with the final aim of a combined
continuum extrapolation of the results obtained in the two cases. 
The first definition of the critical mass is the point where 
the pseudoscalar meson mass, computed with plain Wilson fermions ($\mu =0$) 
vanishes, the second, where the 
PCAC quark mass, computed in the Wilson twisted mass setup, vanishes. 
See refs. \cite{Jansen:2005gf,Jansen:2005gf,Shindler:2005vj}, for details on how the 
critical mass has been computed numerically.
In the following we will refer to the first
situation as the ``pion definition'' and to the second situation 
as the ``PCAC definition'' of the critical point.
Both definitions should lead to $O(a)$-improvement, but 
they can induce very different $O(a^2)$ effects, in particular at 
small pseudoscalar meson masses \cite{Aoki:2004ta,Sharpe:2004ny,Frezzotti:2005gi}. 
Indeed, in refs.~\cite{Bietenholz:2004sa,Bietenholz:2004wv} we reported that 
the pion definition can have substantial $O(a^2)$ effects which are
amplified when the quark mass becomes small and violates the 
inequality \mbox{$\mu > a\Lambda^2$} (where $\mu$ is, at full twist, 
the parameter which provides mass to the pseudoscalar meson). 
On the other hand, when the PCAC definition of the critical
mass is used, these particular kind of $O(a^2)$ cut-off effects are 
dramatically reduced as was shown numerically in refs.~\cite{Abdel-Rehim:2005gz,Jansen:2005gf,Jansen:2005kk}, 
and theoretically
in refs.~\cite{Aoki:2004ta,Sharpe:2004ny,Frezzotti:2005gi}. 

The purpose of the present letter is a demonstration that Wilson 
twisted mass fermions are in a position to reach small quark masses and eventually 
allow a comparison with chiral perturbation theory also for more complicated 
physical observables than the 2-point functions considered in 
refs.~\cite{Jansen:2003ir,Bietenholz:2004sa,Bietenholz:2004wv,Abdel-Rehim:2005gz,Jansen:2005gf,Jansen:2005kk,
Abdel-Rehim:2005qv,Abdel-Rehim:2005yx}.
A computation of pion form factors 
employing the twisted mass fermion approach was already presented in
ref.~\cite{Abdel-Rehim:2004gx}, but at a rather high values of the
pseudoscalar meson mass (470 and 660 MeV), 
using only the pion definition of the critical quark mass.

\subsection{The twist-2, non-singlet operator}

The towers of twist-2 operators related to the unpolarized structure functions
have the following expressions
\be
O^a_{\mu_1 \cdots \mu_N}(x) = \frac{1}{2^{N-1}} \bar\psi(x)\gamma_{\{\mu_1}
\lrD_{\mu_2}\cdots \lrD_{\mu_N\}}\frac{1}{2} \tau^a\psi (x)\; ,
\label{eq:op}
\ee
where $\{\cdots\}$ means symmetrization on the Lorentz indices and
\be
\lrD_{\mu} = \rD_{\mu} - \lD_{\mu}; \qquad  D_{\mu} =
\frac{1}{2}[\nabla_\mu + \nabla_\mu^{*}]\; .
\ee
The flavour structure is specified by the Pauli matrices $\tau^a$ where we
include here also the identity with $\tau^0=2 \cdot \mathds{1}$.   
In general one should perform an axial rotation in order to obtain the
expressions for the twist-2 operators for the twisted mass formulation.
For our purposes it is enough to notice that the operators in
eq. (\ref{eq:op}) with flavour index $a=0,3$ do not rotate.
We concentrate in this
work on the twist-2 quark operator related to the lowest moment 
of the valence quark parton distribution function in a pion.
In particular for the {\it up} quark (the {\it down} quark can be treated in
the same way) this amounts to consider operators of the following form 
\be
{\mathcal O}^u_{\mu \nu}(x) = \frac{1}{2}\bar\psi (x) \gamma_{\{\mu}
\lrD_{\nu\}} \frac{(1+\tau^3)}{2}\psi (x) - \delta_{\mu \nu} \cdot {\rm trace~terms}\; ,
\label{eq:ope1_tr}
\ee

There are
two representations of such a non-singlet operator on the lattice
\cite{Baake:1982ah,Mandula:1983ut}. 
In the following we will concentrate on the operator
\be
\cO^u_{44}(x) = \frac{1}{2} \bar\psi(x) \Big[ \gamma_4 \lrD_4 - {1 \over 3}
\sum_{k=1}^3 \gamma_k \lrD_k \Big] \frac{(1+\tau^3)}{2} \psi(x)\; ,
\label{O44}
\ee
since in computing the matrix elements of this operator one has not to supply
an external momentum (an external momentum increases considerably
the noise to signal ratio).

The matrix elements of this operator can be computed in the standard way,
described in refs. \cite{Martinelli:1987zd,Best:1997qp}. We indicate with 
\be
P^{\pm}(x) = \bar\psi(x) \gamma_5 \frac{\tau^{\pm}}{2} \psi(x)\; , \qquad
\tau^{\pm} = \frac{\tau^1 \pm i\tau^2}{2}
\ee
the interpolating operator for the charged pseudoscalar meson.
The ratio of the 3-point function
\be
C_{44}(y_4) = a^6 \sum_{{\bf x},{\bf y}}\langle P^+(0) \cO_{44}({\bf y},y_4)
P^-({\bf x},T/2) \rangle\; ,
\label{eq:C44}
\ee
and the 2-point function
\be
C_P(x_4) = a^3\sum_{{\bf x}}\langle P^+(0) P^-({\bf x},x_4) \rangle\; ,
\ee
is related to the matrix element we are interested in.
The Wick contractions of the correlation function (\ref{eq:C44}) contain also
a disconnected piece that we neglect consistently with the fact that we are
interested in the valence quark distribution.
In particular if we perform a transfer matrix decomposition and define
\be
R(y_4) = \frac{C_{44}(y_4)}{C_P(T/2)}\; ,
\ee
in the limit when only the fundamental state dominates ($0 \ll y_4
\ll T/2$), and the ratio $R$ reaches a plateau in $y_4$, we obtain
\be
\langle 0,PS| \cO_{44} | 0,PS\rangle = 2 m_{PS} R\; .
\ee
where $| 0,PS\rangle$ indicates the fundamental state in the charged
pseudoscalar channel.
The relevant bare quantity is then 
\be
\langle x \rangle^{\rm bare} = \frac{1}{m_{PS}}\cdot R\; .
\ee
In fig.~{\ref{fig:plateau} we show an example
of such a plateau from which we read off the bare matrix element, from our
second smallest pseudoscalar meson mass and smallest lattice spacing. 

\begin{figure}[htb]
\vspace{-0.0cm}
\begin{center}
\epsfig{file=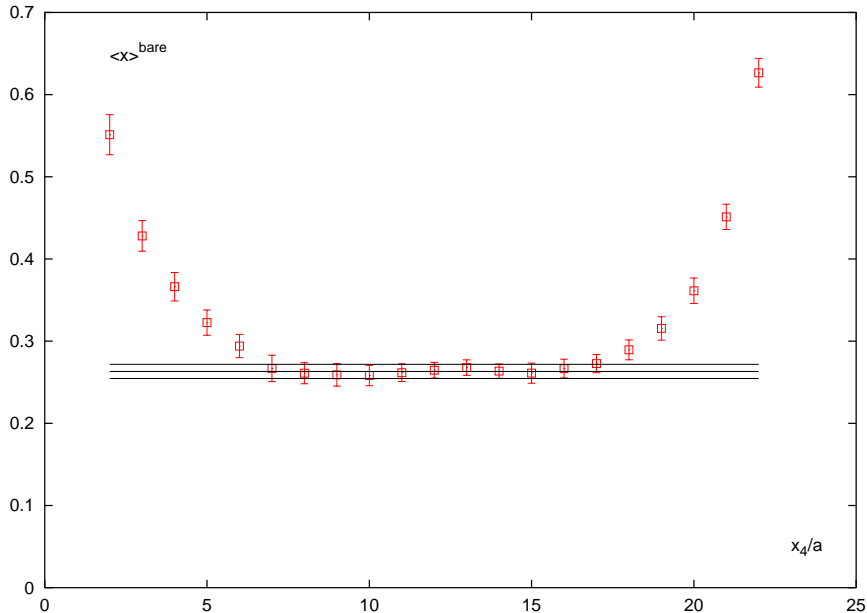,angle=270,width=0.8\linewidth}
\end{center}
\vspace{-0.0cm}
\caption{$\langle x \rangle^{\rm bare}$ at $\beta=6.2$ and a
  pseudoscalar meson mass of
  about 370 MeV; we show the average of the two plateaux around $T/4$ and $3T/4$.
\label{fig:plateau}}
\end{figure}

The matrix element obtained in this way has to be renormalized by a 
multiplicative renormalization factor. 
To this end, we took the $Z_\mathrm{RGI}$ as computed in 
refs.~\cite{Guagnelli:2004ga,Guagnelli:2003hw} from a Schr\"odinger functional
(SF) renormalization scheme \cite{Luscher:1992an,Sint:1993un,Sint:1995rb,Jansen:1995ck},
with standard Wilson action.
The renormalization factor will affect the renormalized matrix element by
$O(a)$ contributions.
In order to check the O($a$) effects coming from the boundaries of the SF 
we have varied the improvement coefficients
typical of the SF boundaries and we have observed no variation in the
continuum limit, indicating that these boundary $O(a)$ effects are negligible.
Nevertheless the continuum limit discussed in the following section is always
performed with the boundary improvement coefficients set to their perturbative values.

The renormalized matrix element has a well defined continuum limit and in the
phenomenologically relevant $\msbar$ scheme it is given by 
\be
\langle x \rangle^{\msbar} (\mu,r_0m_{PS}) = \lim_{a\to 0}
\frac{\langle x \rangle^{\rm bare}(a,m_{PS}) Z_\mathrm{RGI}(a) }{f^{\msbar}(\mu)} \; \qquad
\mu=2~ {\rm GeV} ,
\ee
where $Z_\mathrm{RGI}$ and $f^{\msbar}(\mu)$ were computed in
ref. \cite{Guagnelli:2004ga} (see this reference for further details).
We remind here that this $\mu= 2$ GeV indicates the renormalization scale and
not the twisted mass, since we are using a mass independent renormalization scale.

\section{Numerical results}

Our quenched simulations were performed for a number of bare quark masses in a
corresponding pseudoscalar meson mass range of $270 \mathrm{~MeV} < m_{\rm PS} < 1.2
\mathrm{~GeV}$ using the Wilson plaquette gauge action, employing 
periodic boundary conditions for all fields.
In table~\ref{table:simpara} we give further details
of our simulation parameters.

\begin{table}[!t]
\begin{center}
\begin{tabular}{|c||c|c|c|c|c|}
\hline
\hline
$\beta$ &  5.85 & 6.00 & 6.10 & 6.20  & 6.45 \\
\hline   
$a$ (fm) & 0.123 & 0.093 & 0.079 & 0.068 & 0.048 \\
$r_0/a$ & 4.067 & 5.368 & 6.324 & 7.360 & 10.458\\
$L/a$ & 16 & 16 & 20 & 24 & 32\\
$T/a$ & 32 & 32 & 40 & 48 & 64\\
\hline
\hline
&\multicolumn{5}{|c|}{pion definition ($\kcpi$)}\\
\hline
$N_{\rm meas}$ & 255 & 388 & 300 & 207 & 214 \\
\hline
$\mu_2 a$& 0.0100 & 0.0076 & 0.0064 & 0.0055 & 0.0039\\
$\mu_3 a$& 0.0200 & 0.0151 & 0.0128 & 0.0111 &       \\
$\mu_4 a$& 0.0400 & 0.0302 & 0.0257 & 0.0221 &       \\
$\mu_5 a$& 0.0600 & 0.0454 & 0.0385 & 0.0332 &       \\
$\mu_6 a$& 0.0800 & 0.0605 & 0.0514 & 0.0442 &       \\
$\mu_7 a$& 0.1000 & 0.0756 & 0.0642 & 0.0553 &       \\
\hline                                                     
\hline                                                     
&\multicolumn{5}{|c|}{PCAC definition ($\kcP$)}\\
\hline
$N_{\rm meas}$ & 400 & 300 & & 300 &  \\
\hline
$\mu_1 a$&  0.0050 & 0.0038 &   & 0.0028 &\\
$\mu_2 a$&  0.0100 & 0.0076 &   & 0.0055 &\\
$\mu_3 a$&  0.0200 & 0.0151 &   & 0.0111 &\\
$\mu_4 a$&  0.0400 & 0.0302 &   & 0.0221 &\\
$\mu_5 a$&  0.0600 & 0.0454 &   & 0.0332 &\\
$\mu_6 a$&  0.0800 & 0.0605 &   & 0.0442 &\\
$\mu_7 a$&  0.1000 & 0.0756 &   & 0.0553 &\\\hline
\hline
\end{tabular}
\end{center}
\caption{\it Simulation parameters and the number of measurements ($N_{\rm meas}$)}
\label{table:simpara}
\end{table}

\begin{table}[!t]
\begin{center}
\begin{tabular}{|c||c|c|c|c|c|c|}
\hline
\hline
$\beta$ &  5.85 & 6.00 & 6.10 & 6.20  & 6.45 \\   
\hline
\hline
&\multicolumn{5}{|c|}{$m_{\rm PS} a$ ($\kcpi$)}\\
\hline
$\mu_2 a$& 0.2240(33) & 0.1773(43)&  0.1482(27) &   0.1282(24)&0.0892(22)\\
$\mu_3 a$& 0.3117(23) & 0.2379(29)&  0.2030(21) &   0.1760(20)&\\
$\mu_4 a$& 0.4430(33) & 0.3337(22)&  0.2865(15) &   0.2472(16)&\\
$\mu_5 a$& 0.5523(23) & 0.4135(17)&  0.3534(13) &   0.3055(14)&\\
$\mu_6 a$& 0.6478(21) & 0.4840(16)&  0.4130(13) &   0.3561(12)&\\
$\mu_7 a$& 0.7349(21) & 0.5491(14)&  0.4676(12) &   0.4034(11)&\\
\hline                                                     
\hline                                                     
&\multicolumn{5}{|c|}{$m_{\rm PS} a$ ($\kcP$)}\\
\hline
$\mu_1 a$&  0.1640(24) &   0.1178(69) &   &  0.0936(22)  &\\
$\mu_2 a$&  0.2289(19) &   0.1686(50) &   &  0.1278(20)  &\\
$\mu_3 a$&  0.3231(14) &   0.2401(33) &   &  0.1781(17)  &\\
$\mu_4 a$&  0.4608(12) &   0.3422(23) &   &  0.2495(13)  &\\
$\mu_5 a$&  0.5703(11) &   0.4233(18) &   &  0.3080(11)  &\\
$\mu_6 a$&  0.6659(10) &   0.4943(15) &   &  0.3588(10)  &\\
$\mu_7 a$&  0.7532(9)  &   0.5595(15) &   &  0.4062(9)   &\\
\hline
\hline
\end{tabular}
\end{center}
\caption{\it Pseudoscalar meson masses $m_{\rm PS} a$ for all simulation
  points. These data refer to a subset of the data obtained in ref.~\cite{Jansen:2005kk}.
}
\label{table:mpi}
\end{table}

\begin{table}[!t]
\begin{center}
\begin{tabular}{|c||c|c|c|c|c|c|}
\hline
\hline
$\beta$ &  5.85 & 6.00 & 6.10 & 6.20  & 6.45 \\   
\hline
\hline
$Z^{\msbar}$ & 0.90(5) & 0.95(4) & 0.99(4) & 1.01(4) & 1.06(5) \\   
\hline
\hline
&\multicolumn{5}{|c|}{$\langle x \rangle^{\rm bare,SF}$ ($\kcpi$)}\\
\hline
$\mu_2 a$&  0.2848(105) & 0.2143(107)& 0.2028(93) & 0.2073(103)&0.2226(108)\\
$\mu_3 a$&  0.3295(57)  & 0.2818(56) & 0.2700(53) & 0.2621(49) &\\
$\mu_4 a$&  0.3643(37)  & 0.3294(33) & 0.3144(28) & 0.2996(33) &\\
$\mu_5 a$&  0.3840(25)  & 0.3533(24) & 0.3374(21) & 0.3215(26) &\\
$\mu_6 a$&  0.4012(20)  & 0.3711(20) & 0.3551(17) & 0.3386(24) &\\
$\mu_7 a$&  0.4168(18)  & 0.3861(17) & 0.3700(15) & 0.3539(20) &\\
\hline                                                   
\hline                                                     
&\multicolumn{5}{|c|}{$\langle x \rangle^{\rm bare}$ ($\kcP$)}\\
\hline
$\mu_1 a$&0.2566(157) & 0.2615(219) &   & 0.2505(241)&\\
$\mu_2 a$&0.3049(80)  & 0.2819(110) &   & 0.2698(98) &\\
$\mu_3 a$&0.3420(39)  & 0.3135(58)  &   & 0.2907(46) &\\
$\mu_4 a$&0.3704(24)  & 0.3439(36)  &   & 0.3127(31) &\\
$\mu_5 a$&0.3900(18)  & 0.3631(26)  &   & 0.3300(24) &\\
$\mu_6 a$&0.4077(15)  & 0.3791(21)  &   & 0.3455(19) &\\
$\mu_7 a$&0.4237(13)  & 0.3935(18)  &   & 0.3502(16) &\\
\hline
\hline
\end{tabular}
\end{center}
\caption{\it Renormalization factor $Z^{\msbar}\equiv Z_\mathrm{RGI}(a)/f^{\msbar}(\mu)$ for
  $\mu=2$ {\rm GeV} from ref.~\cite{Guagnelli:2004ga} and bare matrix element
$\langle x \rangle^{\rm bare}$ for all simulation points.}
\label{table:xbare}
\end{table}

\begin{table}[!t]
\begin{center}
\begin{tabular}{|c||c|c|}
\hline
\hline
$m_{\rm PS}$ [GeV] & $\langle x \rangle^{\msbar}$\\
\hline
\hline
$0.272^*$ &$0.260(31)^*$\\
0.368 & 0.243(21)\\
0.514 & 0.272(21)\\
0.728 & 0.299(22)\\
0.900 &0.317(23)\\
1.051 &0.335(24)\\
1.163 &0.350(25)\\
\hline
\hline
\end{tabular}
\end{center}
\caption{\it $\langle x \rangle^{\msbar}$ in the continuum using a combined
  extrapolation of data obtained with the PCAC and pion definition of 
  $\kappa_c$.$^*$The value corresponding to a pseudoscalar meson mass of 
272 MeV has not been corrected for FSE.} 
\label{table:cont}
\end{table}

We performed simulations at different values of the twisted mass parameter 
$\mu$ while setting $m_0$ to its critical value as obtained
from the pion or the PCAC definition. The corresponding critical hopping
parameters can be found in ref.~\cite{Jansen:2005kk}. The results are
summarized in table \ref{table:mpi} for the pseudoscalar mass and
in table \ref{table:xbare} for the bare matrix element.

The goal of this letter is to perform the continuum extrapolation of 
$\langle x \rangle$ at a fixed value of $m_{\rm PS} r_0$, for a number of
values of $m_{\rm PS} r_0$. An interpolation of the values of 
$\langle x \rangle$ to the chosen values of $m_{\rm PS} r_0$ is needed. 
These values are close to the simulated ones, and so even a linear
interpolation is usually sufficient. By using the value of the force parameter 
$r_0=0.5 \rm{~fm}$ \cite{Sommer:1993ce,Guagnelli:1998ud} to set the
scale, the lowest pseudoscalar meson mass that can be reached corresponds to 
$m_{\rm PS}=272$ MeV (for which we have only data obtained with the PCAC
definition of $\kappa_c$). On the basis of the study performed in
ref.~\cite{Guagnelli:2004ww}, we expect finite size effects (FSE) in the matrix element 
to be relevant for the smallest four quark masses
simulated. Extending the study of ref.~\cite{Guagnelli:2004ww} down to values of 
$m_{\rm PS}L\simeq 2.7$ (for which the FSE can be as large as $13\%$), 
we have corrected the matrix elements for these effects down to 
the second smallest quark mass ($m_{\rm PS}$=368 MeV). For the
smallest quark mass, however, the sensitivity required to investigate FSE 
is computationally very expensive and here 
we present the corresponding points without corrections, with the purpose of
showing that, even for quantities more complicated to extract than meson 
masses or decay constants, there are no problems of principle in reaching 
small quark masses.  

In fig.~\ref{fig:me} we show the combined continuum extrapolation of 
$\langle x\rangle$ obtained with the two definitions of $\kappa_c$, 
already converted to the $\msbar$ scheme at $\mu=2$ GeV as explained
in the previous section, for a wide range of values of fixed pseudoscalar meson masses.
In principle, employing the renormalization factors obtained
with (untwisted) Wilson fermions O($a$) lattice
artifacts can be introduced, which are absent in the bare matrix elements. However this
kind of O($a$) effects are independent of the mass and of the definition
of the $\kappa_c$ used. Considering for example the case of the second 
lowest mass $\mu_2$ ($m_{PS}=368$ GeV) in fig.~\ref{fig:me} 
and performing a combined continuum fit of the type $A+B a/r_0 + C
(a/r_0)^2$ where $A$ and $B$ are the same for the two definitions of
$\kappa_c$ while $C$ is different, one finds that $B\approx 0$.
Since $B/A$ does not depend upon the mass it follows that in the 
determination of the renormalization factors O($a$) lattice artifacts
are in practice negligible. We are thus justified in performing a
continuum extrapolation of the type $A+ C(a/r_0)^2$ and we can see from  
fig.~\ref{fig:me} that the scaling of $\langle x\rangle$ is in agreement 
with pure O($a^2$) cut-off effects for both definitions of the critical mass
and values of $\beta\geq 6.0$. 
The slope of $\langle x\rangle$ as a function of $a^2$ appears to be
rather small for the PCAC definition of $\kappa_c$.
At $\beta=5.85$ and lower, lattice artifacts which increase with 
$\mu$ are visible for the highest masses while, in analogy to what 
observed in ref.~\cite{Jansen:2005kk}, by using the pion
definition of $\kappa_c$ O($a^2$) cut-off effects are enhanced at small
quark mass. Indeed, for the second 
smallest pseudoscalar meson mass (the smallest one is absent with the pion
definition of $\kappa_c$), we performed an additional simulation at
$\beta=6.45$,
in order to have a better control on the
continuum extrapolation, 
and we excluded the point at $\beta=6.0$ (which appears to be outside of the
scaling region). 
Fig.~\ref{fig:me} nicely demonstrates that using
only this definition of the critical mass it is important to add the data
point at $\beta=6.45$ and this effect would be probably even worse at
the smallest pseudoscalar meson mass. 

\begin{figure}[htb]
\vspace{-0.0cm}
\begin{center}
\epsfig{file=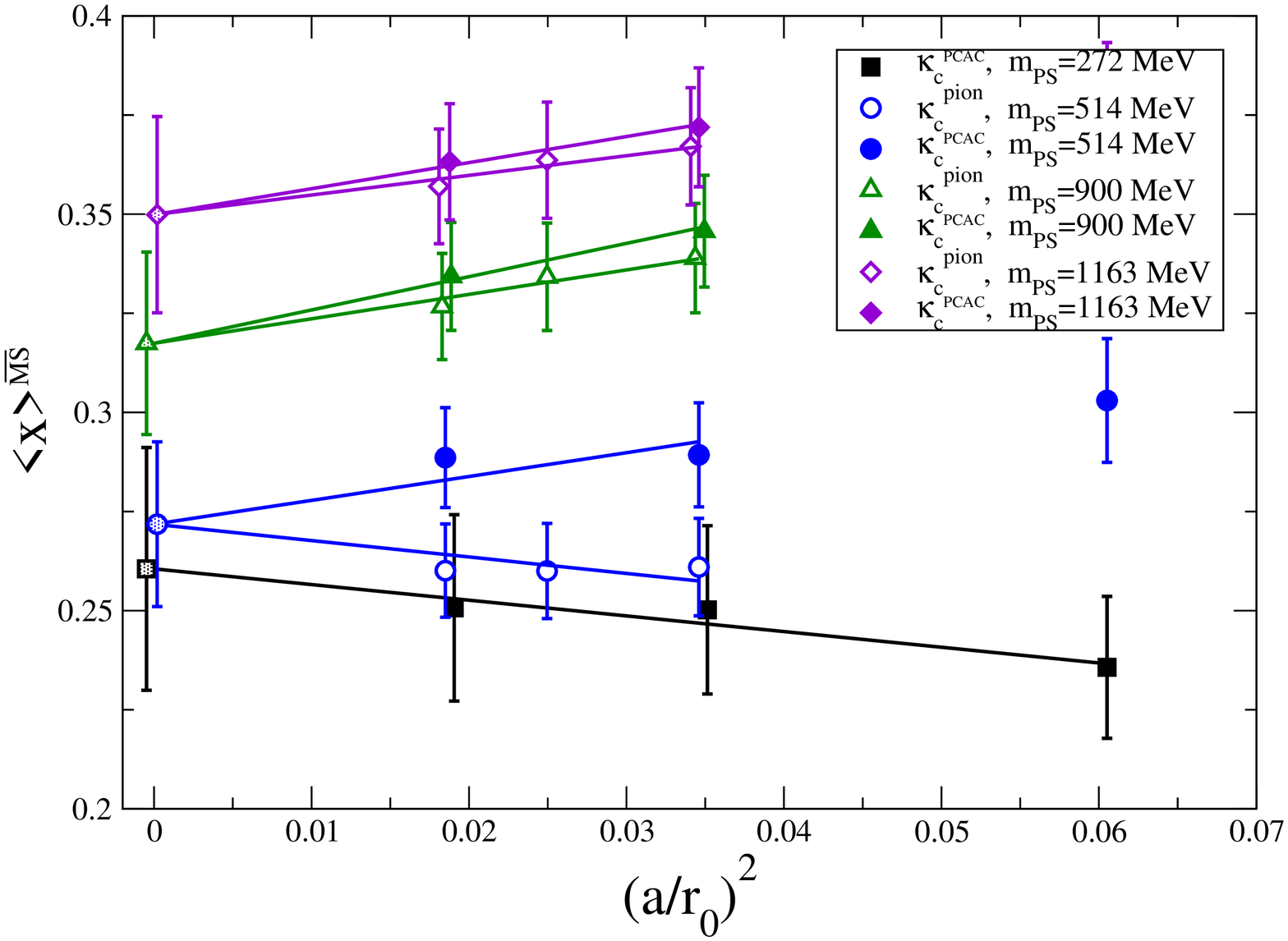,angle=270,width=0.8\linewidth}
\epsfig{file=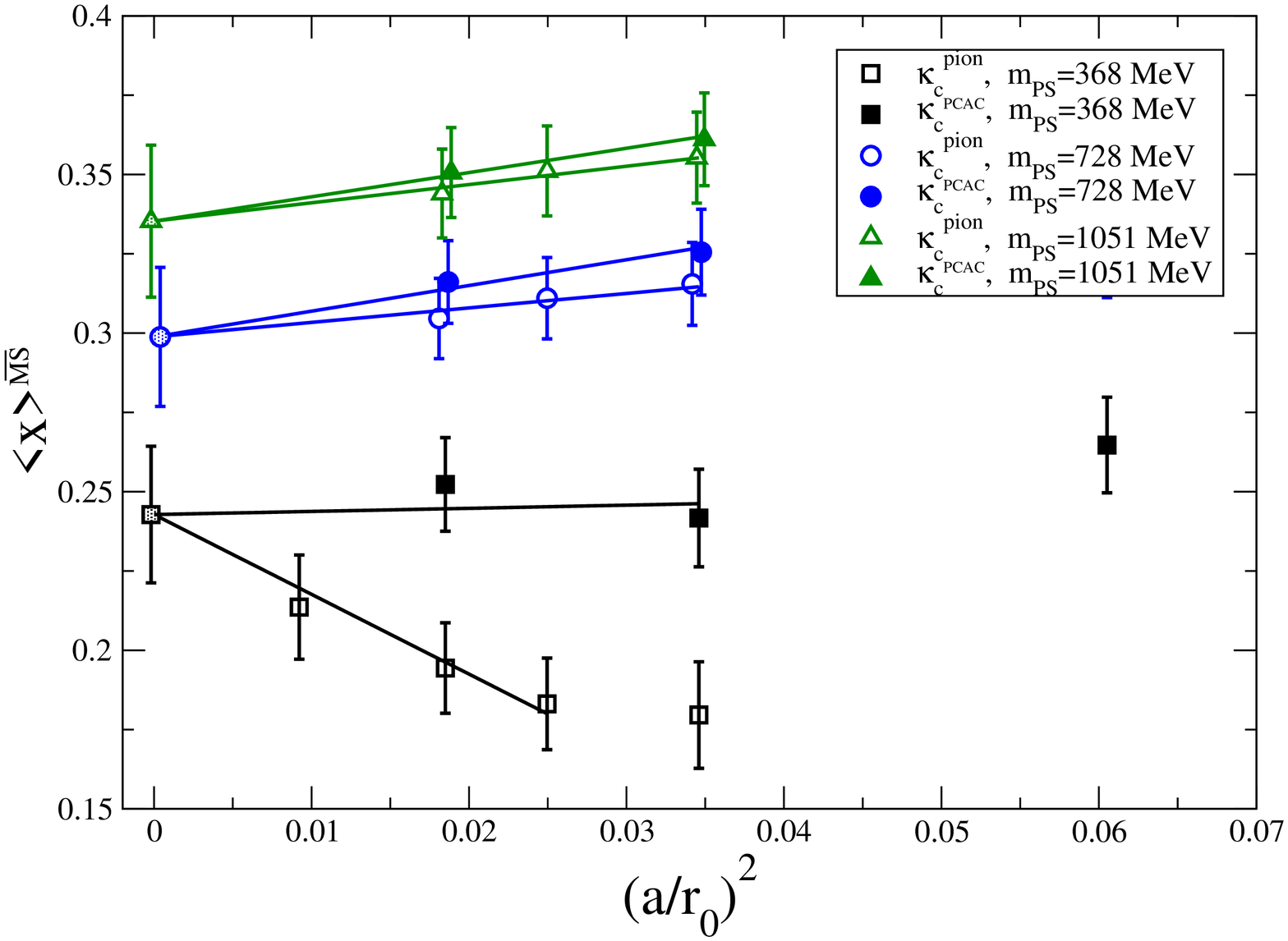,angle=270,width=0.8\linewidth}
\end{center}
\vspace{-0.0cm}
\caption{$\langle x \rangle^{\msbar}$ as a function of $(a/r_0)^2$ for different
  values of the pseudoscalar meson mass.  
\label{fig:me}}
\end{figure}

In fig.~\ref{fig:continuum} and table \ref{table:cont} we present the
results for 
$\langle x \rangle^{\msbar}(\mu=2 {\rm GeV})$ in the continuum as a 
function of the pseudoscalar meson mass in GeV. The empty squares are our
values obtained earlier from a combined continuum extrapolation of Wilson
and clover-improved Wilson fermions data using the Schr\"odinger functional scheme
\cite{Guagnelli:2004ga}. As usual, such quenched simulations have to stop at a
pseudoscalar meson mass
of about 600 MeV. At such high masses it becomes very difficult, if not
impossible, to compare the simulation results to chiral perturbation theory
\cite{Bernard:2003rp,Chen:2001gr,Beane:2003xv} or to other phenomenological
predictions \cite{Leinweber:1998ej,Detmold:2001jb}, even when the results are
extrapolated to the continuum limit as done here. 

Fig.~\ref{fig:continuum} shows that 
with Wilson twisted mass fermions, the large gap between pseudoscalar
meson masses of 
about 600 MeV, as the lower bound for standard simulations, and the physical
value can be bridged. Quenched chiral perturbation theory predicts
\cite{Chen_priv} the absence of chiral logs for the matrix element studied in this letter,
but our present large error bars, and the lack of more data in the region ($m_{\rm PS} \lesssim 500$ MeV)
where chiral perturbation theory should be applicable, does not allow us to perform a 
careful chiral extrapolation. We quote then as our final result 
\be
\langle x \rangle^{\msbar} (2 {\rm GeV}) = 0.243(21)
\ee
given by the value at the next to smallest pion mass.

\begin{figure}[htb]
\begin{minipage}{\linewidth}
%\vspace{-0.0cm}
\begin{center}
\epsfig{file=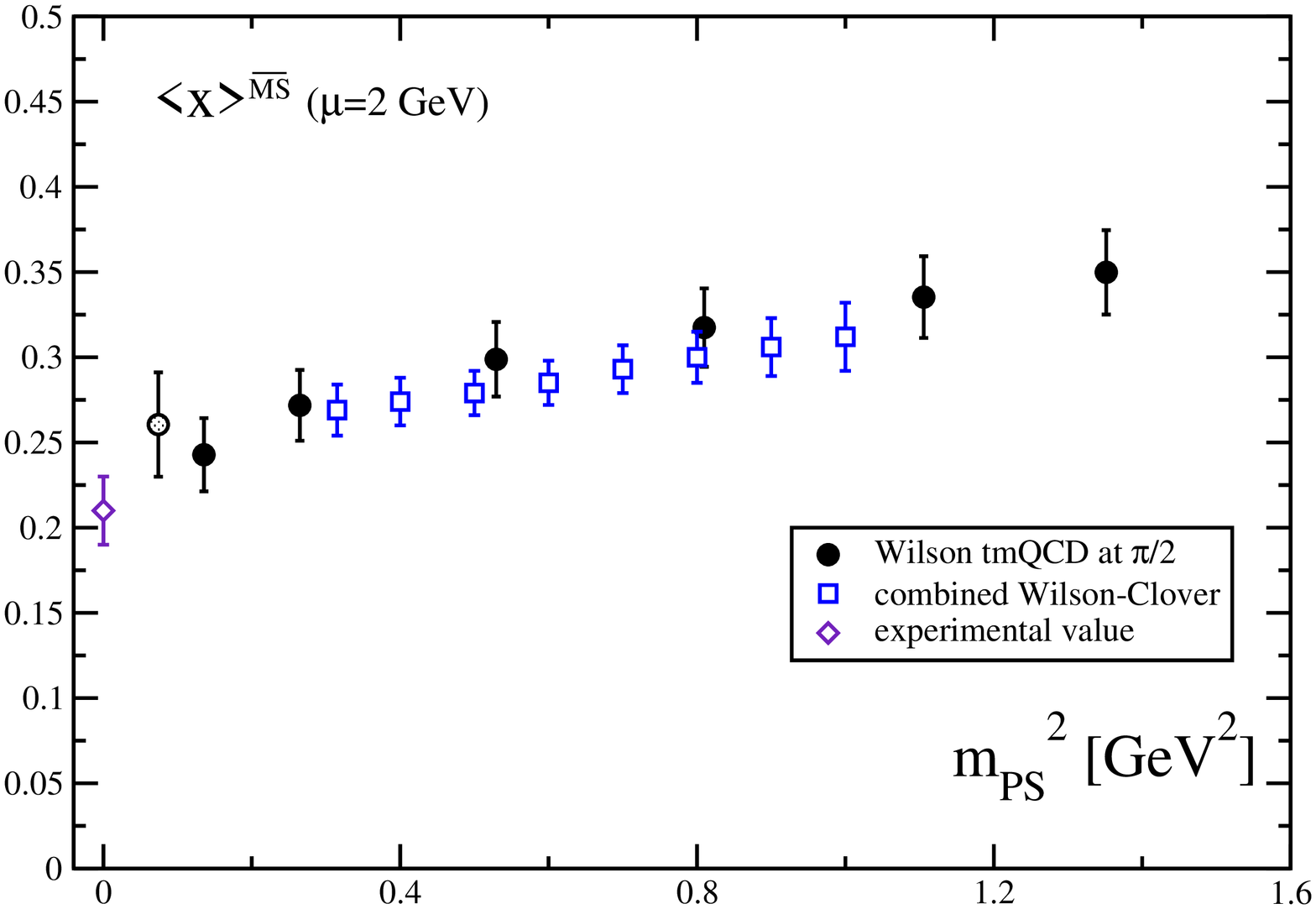,angle=270,width=0.8\linewidth}
\end{center}
%\vspace{-0.0cm}
\caption[For a LoF]{$\langle x \rangle^{\msbar}(\mu= \ 2 \ {\rm GeV})$ extrapolated to the continuum as a function of 
the pseudoscalar meson mass. Squares are obtained from a combined continuum 
extrapolation of earlier Wilson and clover improved Wilson simulations 
{\protect\cite{Guagnelli:2004ga}}. The circles represent our results using 
Wilson twisted mass fermions. For the empty circle at the smallest mass see
the text. The diamond represents the experimental
value as obtained from global fits \protect{\cite{Sutton:1991ay,Gluck:1999xe}}. 
Recently a new analysis {\protect\cite{Wijesooriya:2005ir}} has been performed giving as a result 
$\langle x \rangle^{\msbar}(\mu= \ 2.28 \ {\rm GeV}) = 0.217(11)$.
\footnote{We thank C. Roberts to bring this reference to our attention.}}
\label{fig:continuum}
\end{minipage}
\end{figure}

\section{Conclusions}

In this letter we performed a test of Wilson twisted mass fermions for 
interesting physical observables, the moments of parton distribution 
function which are relevant in deep inelastic scattering. 
So far, most investigations
of Wilson twisted mass QCD at small masses considered only 2-point 
correlation functions. The present study is the first that investigates 
3-point correlators down to masses of order 250 MeV.
In particular, here we have studied the example of a twist-2 operator. 
The matrix element of such a renormalized operator between pion states 
$\langle x\rangle$ corresponds to the average momentum of the valence quark
distribution ({\it up} for example) in a pion.
 
In the present work we employed two definitions of the critical
mass, the pion and PCAC definition \cite{Jansen:2005gf,Jansen:2005kk}. 
The scaling of $\langle x\rangle$ is in agreement with the expected 
O($a^2$) cut-off effects for both definitions of the critical mass.
We could perform a controlled continuum extrapolation of $\langle x\rangle$
down to pseudoscalar meson masses of about 270 MeV by combining the 
data obtained
with the two definitions.  
Of course, in principle, also overlap simulations are able to reach such
values of the pseudoscalar meson mass. 
This will come, however, at a much higher 
simulation cost \cite{Chiarappa:2004ry}.
Our final figure, fig.~\ref{fig:continuum}, clearly demonstrates that with our
present setup it is possible to 
bridge the gap between large pseudoscalar meson mass values of 600
MeV and the physical value of the pion mass.

\section{Acknowledgements}
We thank Stefan Sint for discussions. 
The computer centers at NIC/DESY Zeuthen, NIC at Forschungszentrum
J{\"u}lich and HLRN provided the necessary technical help and computer
resources. 
S.~C.~gratefully acknowledges support by Fonds zur F\"orde-\\rung der
Wissenschaftlichen Forschung in \"Osterreich, Project P16310-N08.
This work was supported by the DFG 
Sonderforschungsbereich/Transregio SFB/TR9-03.
This work has been also supported in part by the EU Integrated
Infrastructure Initiative Hadron Physics (I3HP) under contract
RII3-CT-2004-506078.

\bibliographystyle{JHEP}
\bibliography{3pt}

\end{document}